\documentclass[
reprint,
superscriptaddress,
tightenlines,
nofootinbib,
amsmath,amssymb,
aps,
pra,
longbibliography
]{revtex4-2}

\usepackage{amssymb}
\usepackage{scrextend}
\usepackage{placeins}
\usepackage{appendix}
\usepackage{graphicx}
\usepackage{dcolumn}
\usepackage{bm}
\usepackage{booktabs}
\usepackage{hyperref}
\usepackage{xcolor}
\hypersetup{
    colorlinks,
    linkcolor={blue!70!black},
    citecolor={blue!70!black},
    urlcolor={blue!70!black}
}

\newcommand{\Xstate}{\tilde{X}\,\,{}^2 \! A_1}
\newcommand{\Astate}{\tilde{A}\,\,{}^2 \! E}
\newcommand{\wavenumber}{cm$^{-1}$}

\DeclareMathOperator*{\dprime}{\prime \prime}

\begin{document}
 
\title{Observation and laser spectroscopy of ytterbium monomethoxide, YbOCH$_{3}$}

\author{Benjamin L. Augenbraun}
\email{augenbraun@g.harvard.edu}
\author{Zack D. Lasner}
\author{Alexander Frenett}
\author{Hiromitsu Sawaoka}
\affiliation{Department of Physics, Harvard University, Cambridge, MA 02138, USA}
\affiliation{Harvard-MIT Center for Ultracold Atoms, Cambridge, MA 02138, USA}

\author{Anh T. Le}
\affiliation{School of Molecular Science, Arizona State University, Tempe, AZ 85287, USA}

\author{John M. Doyle}
\affiliation{Department of Physics, Harvard University, Cambridge, MA 02138, USA}
\affiliation{Harvard-MIT Center for Ultracold Atoms, Cambridge, MA 02138, USA}

\author{Timothy C. Steimle}
\email{TSteimle@asu.edu}
\affiliation{School of Molecular Science, Arizona State University, Tempe, AZ 85287, USA}

\date{December 2, 2020}

\begin{abstract}
\noindent We describe a laser spectroscopic study of ytterbium monomethoxide, YbOCH$_3$, a species of interest to searches for time-reversal symmetry violation using laser-cooled molecules. We report measurements of vibrational structure in the $\tilde{X}$ and $\tilde{A}$ states, vibrational branching ratios for several components of the $\tilde{A}$ state, and radiative lifetimes of low-lying electronic states. \textit{Ab initio} calculations are used to aid the assignment of vibronic emission bands and provide insight into the electronic and vibrational structure. Our results demonstrate that rapid optical cycling is feasible for YbOCH$_3$, opening a path to orders-of-magnitude increased sensitivity in future measurements of $P$- and/or $T$-violating physics.
\end{abstract}

\maketitle

\section{Introduction}
Precision measurements of symmetry-violating electromagnetic moments provide stringent tests of physics beyond the Standard Model (BSM)~\cite{Bernreuther1991, Pospelov2005, Engel2013, DeMille2015, Nakai2017, Cesarotti2019}. Polar molecules containing a high-$Z$ nucleus have emerged as ideal probes because they may combine intrinsic sensitivity to CP-violating moments with structural features that aid high-precision experiments~\cite{DeMille2015, Meyer2006, Meyer2008, Hutzler2020, Hudson2011}. For example, experiments using a variety of (diatomic) polar molecules have set increasingly tight limits on the value of an electron electric dipole moment (eEDM), ultimately constraining CP-violating BSM physics at energy scales $\gtrsim$10~TeV~\cite{Hudson2011, ACME2014, ACME2018, Cairncross2017}. Molecules with enhanced sensitivity to a nuclear magnetic quadrupole moment (nMQM) have been proposed for similar tests~\cite{Maison2019, Denis2020MQM}. Concurrent with these advances in precision measurement, molecular laser cooling has recently been extended to both diatomic~\cite{Lim2018} and triatomic~\cite{Augenbraun2020YbOH} species containing the heavy Yb nucleus. Measurements combining properly chosen molecular species with laser cooling to the $\mu$K regime are predicted to provide several orders of magnitude increased sensitivity to both the eEDM and the nMQM, potentially allowing experiments to probe PeV-scale BSM physics~\cite{Kozyryev2017PolyEDM}. 

A number of laser-coolable molecules have been proposed for next-generation eEDM measurements, with linear species garnering the most attention due to their structural simplicity~\cite{Kozyryev2017PolyEDM, Lim2018, Isaev2017, Gaul2019, Denis2019, Nakhate2019}. The degenerate vibrational bending mode in YbOH is a promising venue for future eEDM or nMQM measurements due to its small parity doublet, analogous to the $\Omega$-doublets that have allowed strong systematic error rejection in eEDM experiments~\cite{Meyer2008, Eckel2013, ACME2014, ACME2018, Cairncross2017}. The recent laser cooling of YbOH to $\ll$1~mK in one dimension represents an important first step toward eEDM measurements with laser-cooled polyatomic molecules~\cite{Augenbraun2020YbOH}.  
However, in YbOH the energy of the first bending excitation is relatively high ($\omega_2 \sim 330$~\wavenumber{}) and  anharmonic~\cite{Mengesha2020}, limiting feasible coherence times to $\lesssim 800$~ms~\cite{LanChangPrivate}. 

By contrast, nonlinear symmetric top molecules possess long-lived parity-doubled states due to rigid body rotation about the molecular symmetry axis (``$K$-doublets").\footnote{The quantum number $K$ denotes the molecule-frame projection of $N$, the total angular momentum excluding spin. States with $\lvert K \rvert > 0$ are doubly degenerate. Note that, necessarily, $N \geq \lvert K \rvert$.} For instance, the $K^{\dprime} = 1$ level in YbOCH$_3$ is expected to have a spontaneous lifetime much longer than 100~s due to its low energy ($\sim$10~\wavenumber{}) and proton spin statistics. This could lead to $> 30 \times$ improved statistical sensitivity to BSM physics if full advantage were taken of the feasible coherence times. Moreover, the very small $K$-doubling of metastable rotational levels enables full polarization and internal co-magnetometry, two features that have enabled robust rejection of systematic errors in recent eEDM measurements~\cite{Eckel2013, ACMELongPaper}. To fully leverage the lifetime of the eEDM-sensitive $K$-doublet in a neutral species requires cooling to ultracold temperatures. Laser cooling of molecules is a proven route toward this end~\cite{Barry2014, truppe2017CaF, Ding2020}. Importantly, the recent 1D laser cooling of CaOCH$_3$ to $<$1~mK shows that the complicated vibrational and rotational
structure in symmetric top molecules does not adversely affect the laser cooling process~\cite{Mitra2020}. Ytterbium monomethoxide (YbOCH$_3$), isoelectronic to CaOCH$_3$, is thus an intriguing candidate for eEDM/nMQM measurements using laser-cooled molecules, but to our knowledge there are no previously reported observations of this species. 

The alkaline-earth monomethoxides, MOCH$_3$ (M = Ca, Sr, Ba), have received considerable spectroscopic attention~\cite{Wormsbecher1982, Crozet2002, Forthomme2011, OBrien1988, Namiki1998, Paul2019, brazier1986laser}. These molecules comprise an alkaline-earth atom ionically and monovalently bonded to a relatively electronegative methoxy ligand~\cite{Ellis2001, Bernath1997}. They have linear M-O-C backbones with three off-axis H atoms ($C_{3v}$ symmetry) in both the ground ($\Xstate$) and low-lying excited ($\Astate$ and $\tilde{B} \, ^2A_1$) states~\cite{brazier1986laser, Crozet2005}. The unpaired valence electron is in a metal-centered, hybrid non-bonding orbital~\cite{Namiki1998} resulting in ``diagonal" vibrational branching ratios (VBRs) and short radiative lifetimes, properties highly favorable for laser cooling~\cite{Augenbraun2020ATM, kozyryev2016MOR}. It is not \textit{a priori} obvious that Yb-containing analogues will share all of these properties, e.g. due to previously observed perturbations in YbF and YbOH~\cite{Lim2017, Mengesha2020}.

Here, we present an initial experimental characterization of the previously unobserved lanthanide-containing symmetric top molecule YbOCH$_3$. This is, to date, the highest-mass neutral symmetric top molecule suitable for precision measurement experiments that has been experimentally characterized. An intense electronic transition near 579~nm has been detected and assigned to the origin band of an $\Astate_{1/2} \leftarrow \tilde{X} \, ^2 A_1$ electronic transition. Dispersed laser-induced fluorescence (DLIF) measurements for YbOCH$_3$ and the isotopologue YbOCD$_3$ are used to characterize the vibrational structure in the ground and excited electronic states. In addition, we have recorded a portion of the high-resolution spectrum of the $0_0^0 \, \Astate_{1/2} \leftarrow \tilde{X} \, ^2 A_1$  band. We complement these measurements with \textit{ab initio} electronic structure calculations and find excellent agreement between theory and experiment. Finally, we discuss the prospects for direct laser cooling of YbOCH$_3$ and its use in future precision measurements.

\section{Experimental Setup}
Molecular beams of YbOCH$_3$ are produced in a setup similar to that used in recent studies of YbOH~\cite{Nakhate2019, Steimle2019, Mengesha2020}. Briefly, a rotating ytterbium rod is ablated at $\sim$20~Hz with a short pulse of 532~nm radiation ($\sim10$~ns, $\sim5$ mJ). The ablation plume is entrained in and reacted with a gas mixture of methanol vapor and Ar in a supersonic expansion. The gas mixture is produced by passing Ar at $\sim 4000$~kPa over room-temperature liquid methanol (CH$_3$OH; vapor pressure $\sim10$~kPa). Typical pulse widths for the molecular beam are $\sim 50$~$\mu$s. We use a number of complementary spectroscopic methods including two-dimensional (2D) spectroscopy, DLIF spectroscopy, radiative decay, and high-resolution excitation spectroscopy to conclusively confirm detection of YbOCH$_3$ and provide initial spectroscopic characterization.

For the 2D spectroscopy~\cite{Reilly2006, Gascooke2011, Kokkin2014} and DLIF measurements, the free-jet expansion is probed approximately 10~cm downstream from the source using radiation from an excimer-pumped, tunable, pulsed dye laser ($\sim$10~ns pulse width, $\sim$3~\wavenumber{} linewidth). The molecular laser-induced fluorescence (LIF) is focused into a  0.67~m, high-efficiency ($f$-number = 6.0) Czerny-Turner-type monochromator with a low-dispersion grating (300~lines/mm). The DLIF from the grating is imaged on a cooled and temporally gated intensified charge-coupled device (ICCD). The CCD array is binned to produce an array of intensities versus emission wavelength. The ICCD can be gated with resolution $<$1~ns, which enables temporal separation of the LIF from background light due to the dye laser and ablation plume. The ICCD gate width is typically set to 200~ns and delayed 10~ns after the pulsed laser. The wavelength calibration and relative sensitivity of the spectrometer were calibrated prior to data collection using an argon pen lamp and blackbody source. 

High-resolution measurements are performed in a separate apparatus, as in Ref.~\cite{Steimle2019}. The molecular beam is produced as described above, but skimmed in order to reduce Doppler broadening. A single-frequency, continuous-wave (cw) dye laser (linewidth $\sim 1$~MHz) is used to excite the molecules, and the resulting fluorescence is detected on a cooled photomultiplier tube (PMT). Laser powers of approximately 20~mW are used, resulting in comparable power- and Doppler-broadened linewidths, typically $\sim0.001$~\wavenumber{}.

\section{Observations}

\subsection{2D Spectroscopy}

\begin{figure*}[t]
\includegraphics[width=2\columnwidth]{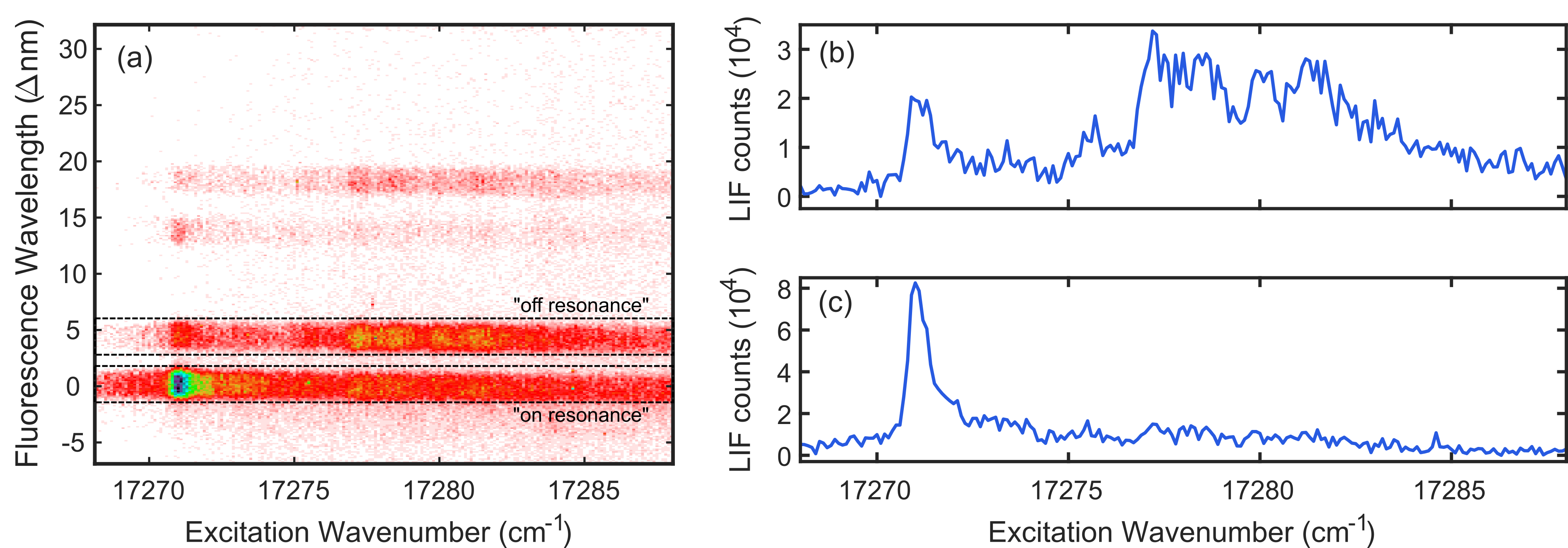} 
\caption{(a) Pulsed-dye laser 2D spectrum in the vicinity of the YbOCH$_3$ $\Astate_{1/2} \leftarrow \Xstate$ bandhead. Integration at fixed fluorescence wavelength can be used to obtain excitation spectra. Color scale indicates number of LIF counts detected. Integration at fixed excitation wavenumber would yield DLIF spectra, but we describe higher resolution DLIF data below. (b) Excitation spectrum obtained by vertical integration in the region labeled ``off-resonance." Strong decay to the bending mode is observed from approximately 10~\wavenumber{} to the blue of the origin frequency (see text). (c) Excitation spectrum by vertical integration in the region labeled ``on-resonance." A strong bandhead associated with the $0_0^0 \, \Astate_{1/2} \leftarrow \Xstate$ vibronic transition is observed, characteristic of a $^2E - ^2A_1$ transition.}
\label{fig:2Dspec}
\end{figure*}

\begin{table}[tb]
\setlength{\tabcolsep}{9pt}
\begin{centering}
\renewcommand{\arraystretch}{1}
\caption{Vibrational normal mode numbering and symmetry labels for MOCH$_3$ molecules under transformations of the $C_{3v}$ point group.}
\begin{tabular}{ccl}
\hline \hline
Mode & Symmetry & Description \\
\hline
 $\nu_1$ & $a_1$ & CH$_3$ symmetric stretch \\
 $\nu_2$ & $a_1$ & CH$_3$ symmetric bend \\
 $\nu_3$ & $a_1$ & O-C stretch \\
 $\nu_4$ & $a_1$ & M-O stretch \\
 $\nu_5$ & $e$ & CH$_3$ asymmetric stretch \\
 $\nu_6$ & $e$ & CH$_3$ asymmetric bend \\
 $\nu_7$ & $e$ & O-CH$_3$ wag \\
 $\nu_8$ & $e$ & M-O-C bend\\

\hline \hline

\end{tabular}
\label{tab:VibLabels}
\end{centering} 
\end{table}

Initial survey scans used 2D (excitation vs. DLIF) spectroscopy to search for YbOCH$_3$ fluorescence. A $75$~nm-wide fluorescence spectral window was monitored while scanning the excitation laser wavelength. To facilitate data analysis, this window tracked the excitation laser as it was scanned. The pulsed dye laser was scanned over a range of about 400~\wavenumber{} near the YbOH $\tilde{A} \, ^2 \Pi_{1/2} \leftarrow \tilde{X} \, ^2 \Sigma^+$ origin band (17323~\wavenumber{}~\cite{Steimle2019}). A strong and broad fluorescence signal was observed around 17271~\wavenumber{}. The 2D spectrum in the vicinity of this signal is shown in Fig.~\ref{fig:2Dspec}(a). The ground state vibrational frequencies observed in emission closely matched those expected of YbOCH$_3$, leading to an initial assignment of this as the YbOCH$_3$ $ 0_0^0 \, \Astate_{1/2} \leftarrow \Xstate$ band. (For convenience, we include the vibrational normal mode numbering and symmetry labeling in Tab.~\ref{tab:VibLabels}.) Ytterbium monohydroxide is also produced in the ablation reaction, and by comparing the intensities of the two species's origin bands we estimate that we produced about an order of magnitude more YbOCH$_3$ than YbOH.

We obtain the excitation spectra shown in Fig.~\ref{fig:2Dspec}(b) by vertically integrating $\pm2$~nm slices of the the off-resonance fluorescence. Similarly, Fig.~\ref{fig:2Dspec}(c) is obtained by vertical integration of the on-resonance fluorescence. Horizontal integration at fixed excitation wavenumber can be used to obtain dispersed fluorescence traces, although these have lower resolution than the DLIF measurements described in Sec.~\ref{sec:ObsDLIF}.


The on-resonance detected excitation spectrum [Fig.~\ref{fig:2Dspec}(c)] exhibits an intense, sharp, blue-degraded band near 17271~\wavenumber{} which is assigned as the $0^0_0 \, \tilde{A}\,^2E_{1/2} \rightarrow \Xstate$ emission. The excitation spectrum is quite compact, implying that the geometry changes little upon excitation and that the electronic orbital angular momentum in the $\tilde{A} \,^2E$ state is largely unquenched (i.e., $\zeta_e \approx 1$)~\cite{Brazier1989, Marr1996}. The off-resonance excitation spectrum, shown in Fig.~\ref{fig:2Dspec}(b), is obtained by vertical integration over a $\pm$2~nm range centered $\sim$130~\wavenumber{} to the red of the excitation wavelength (i.e., by monitoring the Stokes-shifted emission). The extracted excitation spectrum exhibits both the sharp band near 17271~\wavenumber{} and a weaker, broader, and unstructured fluorescence feature near 17281~\wavenumber{}. The latter is assigned as excitation from $\Xstate(v=0)$ to a state of unknown character that we simply designate as [17.28]. The relative intensities of the bands at 17271~\wavenumber{} and 17281~\wavenumber{} for the on- and off-resonance detection suggest that the bending mode in the [17.28] state is quite active. 

Also evident in the 2D spectrum are Stokes-shifted signals centered about 400~\wavenumber{} and 535~\wavenumber{} to the red of the excitation. These are assigned as $4^0_1\, \tilde{A}\,^2E_{1/2}\rightarrow \Xstate$ and $4^0_18^0_1 \Astate_{1/2} \rightarrow \Xstate$, respectively (with similar assignments for the [17.28] bands). Again, the relative intensities of these bands suggest higher bending mode activity in the [17.28] state. It is noteworthy that in YbOH there is also an unassigned vibronic state approximately 10~\wavenumber{} above the $\tilde{A}\,^2\Pi_{1/2}(000)$ state~\cite{Mengesha2020}. A key difference is that in YbOCH$_3$ this higher-energy state couples preferentially to the Yb-O-C bending mode while in YbOH the analogous level exhibits a DLIF spectrum nearly identical to the diagonal origin band.

\begin{figure*}[t]
\centering
\includegraphics[scale=0.50]{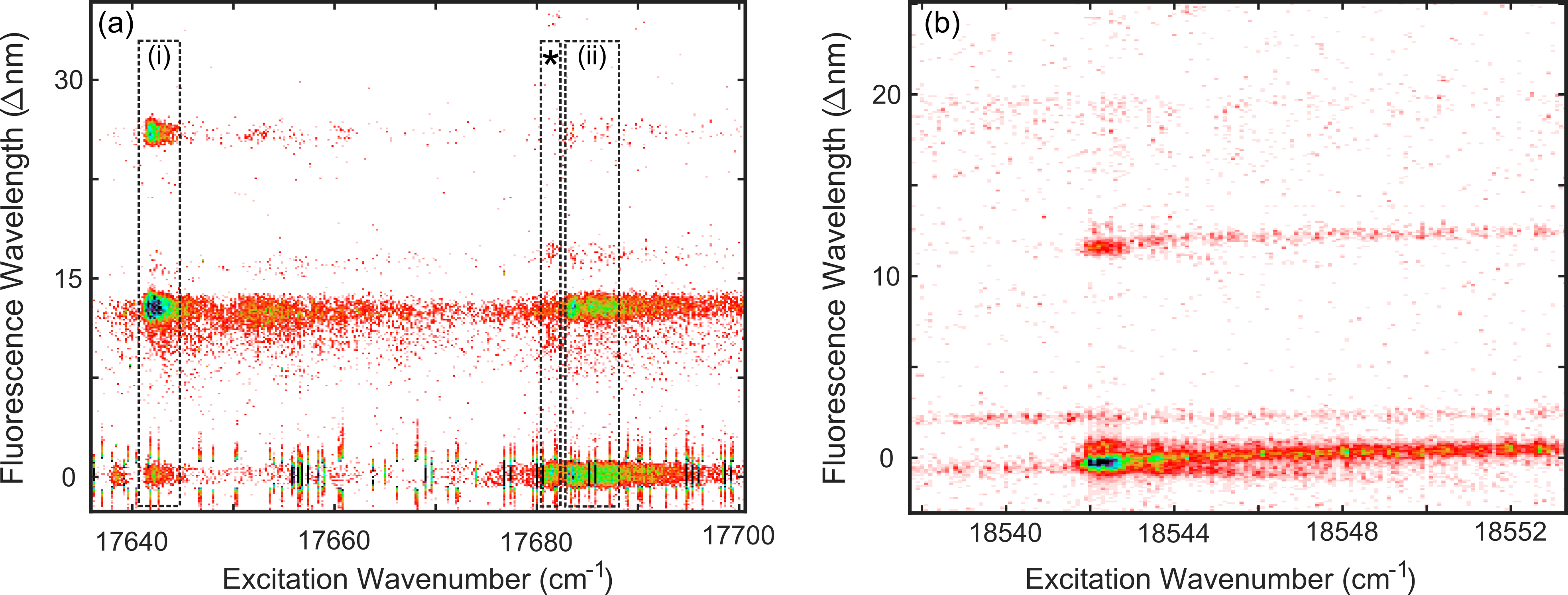} 
\caption{(a) Pulsed-dye laser 2D spectrum in the vicinity of the $4^1_0 \, \Astate_{1/2} \leftarrow \Xstate$ bands. Integration at fixed fluorescence wavelength can be used to obtain excitation spectra. The DLIF features near 17641~\wavenumber{} (marked ``(i)") and 17683~\wavenumber{} (marked ``(ii)") are shown in Fig.~\ref{fig:DLIFnu4}. The weak feature marked by an asterisk is due to the YbOH [17.68] band. (b) Pulsed-dye laser 2D spectrum in the vicinity of the $0^0_0 \, \Astate_{3/2}\leftarrow \Xstate$ band.}
\label{fig:2Dnu4E32}
\end{figure*}

We also recorded 2D spectra in the vicinity of several bands involving excited vibrational levels in the $\Xstate$ and $\Astate$ states. A weak feature was observed about $130$~\wavenumber{} to the red of the origin band near the expected $8^0_1\, \Astate_{1/2} \leftarrow \tilde{X} \, ^2 A_1$ vibronic transition. We observed two bands near the expected position of the $4^1_0 \, \tilde{A}\,^2E_{1/2} \leftarrow \Xstate$ transition, approximately 400~\wavenumber{} above the origin band [Fig.~\ref{fig:2Dnu4E32}(a)]. These bands, near 17641~\wavenumber{} (designated as the [17.64] state) and 17681~\wavenumber{} (designated as the [17.68] state), are blue-shifted by 370~\wavenumber{} and 410~\wavenumber{} relative to the origin band. Based on the long observed progressions in the Yb-O stretching mode, these levels both appear to have strong $v_4=1$ character (see Sec.~\ref{sec:ObsDLIF}). This observation is similar to YbF, where two closely-spaced excited states (the [557] and [561] states) have substantial $v = 1$ character~\cite{Lim2017, Smallman2014}. Fluorescence from these states do not show the strong activity in the $\nu_8$ mode that was observed on the origin band.

In addition, we explored a region approximately 1300~\wavenumber{} above the origin band, where the $\tilde{A}\,^2E_{3/2} \leftarrow \Xstate$ transition would be expected. As shown in Fig.~\ref{fig:2Dnu4E32}(b), a band was observed near 18540~\wavenumber{} with weak off-diagonal decays at frequencies matching the YbOCH$_3$ $\nu_4$ and $\nu_8$ modes. This implies an effective spin-orbit constant of $a \zeta_e d \approx$1270~\wavenumber{}, similar to that of YbF~\cite{Dunfield1995} and YbOH~\cite{SteimleTBD}, indicating that this the spin-orbit interaction is not significantly quenched.

The $\Astate_{1/2}(v=0)$ vibronic state radiative lifetime was measured by fixing the dye laser wavelength to the $0^0_0 \, \Astate_{1/2} \leftarrow \tilde{X} \, ^2 A_1$ bandhead and recording the DLIF spectrum at variable time delay after the pulsed laser excitation. The ICCD gate (200~ns wide) was delayed after the pulsed dye laser in steps of $\sim3$~ns and the resulting fluorescence fit to an exponential to determine the excited state lifetime. The radiative lifetime of the $\tilde{A} \,^2E_{1/2} (v=0)$ level is determined to be 37(4)~ns, somewhat longer than the value of 22(2)~ns for the $\tilde{A}\,^2\Pi_{1/2}(000)$ state of YbOH~\cite{Mengesha2020}.


\subsection{Dispersed Fluorescence}
\label{sec:ObsDLIF}

\begin{figure*}[t]
\centering
\includegraphics[width=2\columnwidth]{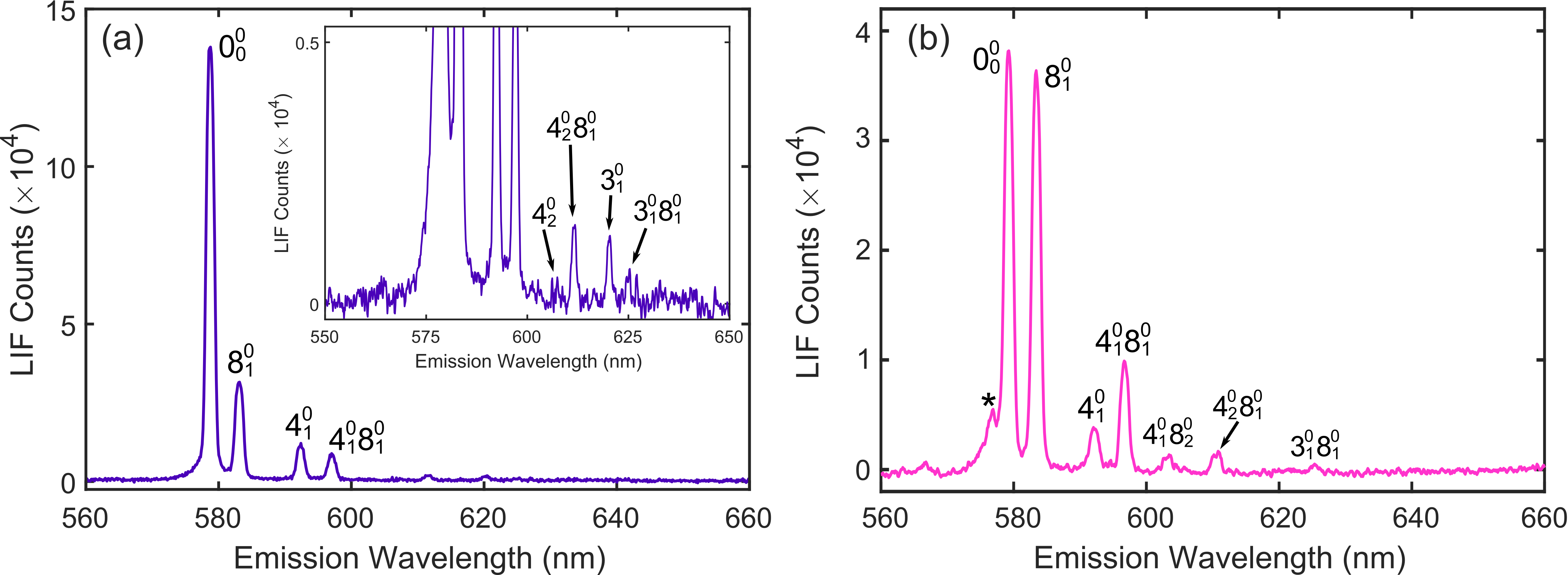} 
\caption{DLIF spectrum resulting from excitation of the  $0^0_0 \, \Astate_{1/2} \leftarrow \tilde{X} \, ^2 A_1$ bandhead. Numbers above each peak indicate transition assignments using the vibrational labeling of Tab.~\ref{tab:VibLabels}. (a) YbOCH$_3$, excitation at 17271.1~\wavenumber{}. Inset: expanded vertical axis to emphasize the weakest decay features observed. (b) YbOCD$_3$, excitation at 17249.7~\wavenumber{}. The small shoulder marked by an asterisk is attributed to impurity YbOH/YbOD.}
\label{fig:DLIF}
\end{figure*}

Higher resolution DLIF spectra were recorded by fixing the excitation laser wavelength and reducing the entrance slit of the monochromator. Typically 5,000 ablation pulses were co-added to achieve sensitive measurement of the vibrational frequencies and FCFs. Background traces were recorded separately to eliminate scattered light due to either the probe or ablation laser.

The DLIF spectrum resulting from pulsed dye laser excitation at the YbOCH$_3$ bandhead ($\sim$17271~\wavenumber{}) is shown in Fig.~\ref{fig:DLIF}(a). The measured vibrational intervals are listed in Tab.~\ref{tab:VibFreqs} and vibrational frequencies are extracted as discussed in Sec.~\ref{sec:DiscXState}. The DLIF is consistent with a relatively diagonal FCF matrix, as expected. A short progression in $\nu_4$ (the Yb-O stretching mode) is observed, with relative intensities similar to those of YbOH~\cite{Mengesha2020}. We observed prominent decay to $\nu_8$ (Yb-O-C bending mode), which was unexpected because the $8^0_1 \, \tilde{A}\,^2E_{1/2} \rightarrow \Xstate$ transition is nominally forbidden by symmetry. Vibronic coupling due to the (pseudo)-Jahn-Teller interaction has previously been invoked to explain similar decays in CaOCH$_3$~\cite{kozyryevCaOCH3}. We do not observe fluorescence at $2\nu_8$, which is symmetry allowed. Weak features associated with the $\nu_3$ (C-O stretch) are also observed, and the frequencies and branching ratios are consistent with values observed in CaOCH$_3$~\cite{kozyryevCaOCH3, Paul2019}.

Isotopic studies aided in the assignment of the ligand-centered vibrations. Figure~\ref{fig:DLIF}(b) displays the DLIF spectrum following excitation in YbOCD$_3$ and vibrational frequencies extracted from these data are listed in Tab.~\ref{tab:VibFreqs}. The origin band of YbOCD$_3$ was found near 17250~\wavenumber{}. This $\approx 20$~\wavenumber{} isotope shift is similar to the large shift ($\approx 10$~\wavenumber{}) observed upon deuteration of CaOCH$_3$~\cite{Crozet2005}. Interestingly, the branching ratios observed for YbOCD$_3$ differ strikingly from those observed in YbOCH$_3$, although the frequencies agree well with the expected isotope shifts. 

\begin{figure*}[tb]
\centering
\includegraphics[width=2\columnwidth]{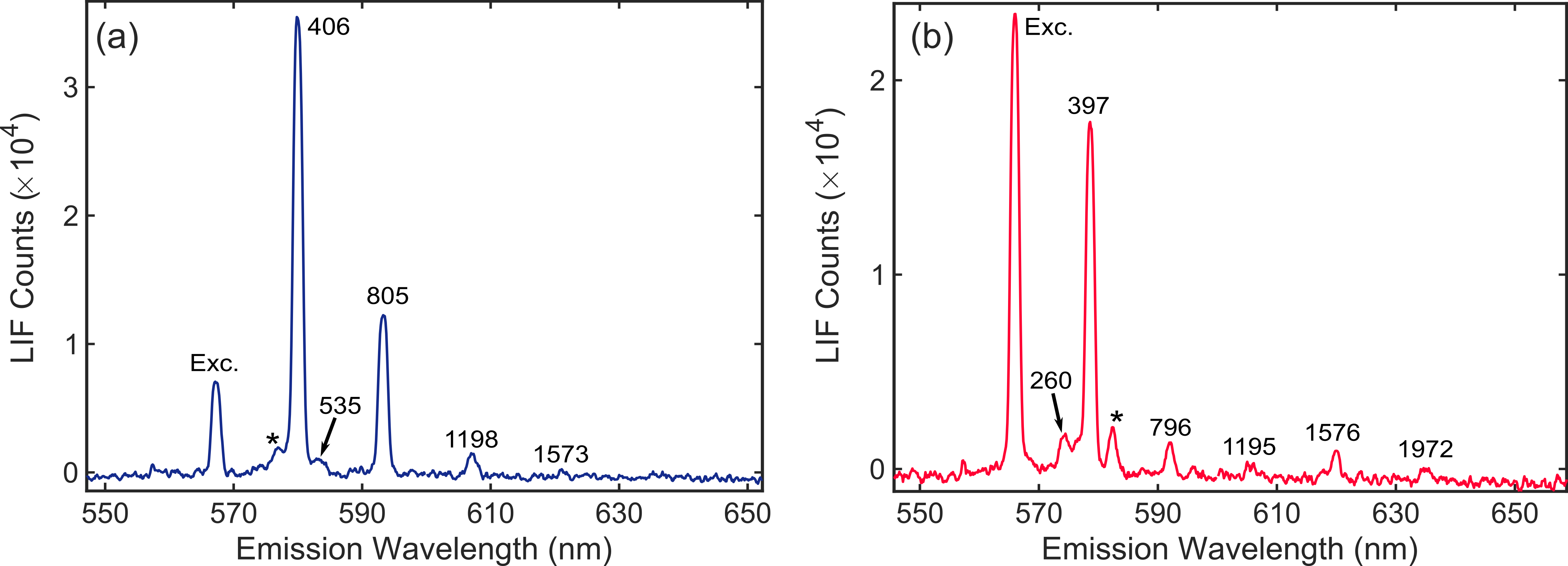} 
\caption{DLIF spectrum resulting from excitation at (a) 17641.75~\wavenumber{} and (b) 17683.40~\wavenumber{}. The numbers above the spectral features are the measured shifts (in cm$^{-1}$) relative to the excitation wavelength (``Exc."). Features marked by an asterisk are due to YbOH impurity fluorescence as described in the text.}
\label{fig:DLIFnu4}
\end{figure*}

We have also measured DLIF spectra following excitation to bands near the expected position of the $\tilde{A}\,^2E_{1/2}(v_4=1)$ state. As described above, two such states were found near 17641~\wavenumber{} and 17683~\wavenumber{}, which we call the [17.64] and [17.68] states, respectively. DLIF measurements from these levels are shown in Figs.~\ref{fig:DLIFnu4}(a) and (b). These spectra both show strong decays to a progression in $\nu_4$, confirming large $v_4=1$ character of each excited state. Weak features associated with decay to $v_8=1$ are also observed, although with significantly lower intensity than observed on the origin band. The presence of two states with significant $v_4=1$ character is likely due to mixing of the $\tilde{A}\,^2E_{1/2} (v_4=1)$ vibronic level with vibronic levels of states associated with $f$-orbital vacancies, e.g. ([Xe]$4f^{13}6s^2$)$_\text{Yb$^+$}$, similar to the case of YbF~\cite{Lim2017, Smallman2014}. This spectral region is complicated by the presence of YbOH bands within a few cm$^{-1}$ of both the [17.64] and [17.68] states. We attribute a few small background features to these YbOH bands, which fluoresce at a range of wavelengths including $\sim$577~nm and $\sim$582~nm~\cite{Mengesha2020}.

\begin{table*}[t]
\setlength{\tabcolsep}{9pt}
\begin{centering}
\renewcommand{\arraystretch}{1}
\caption{Observed and calculated (\textit{ab initio}) vibrational intervals in $\Xstate$ for YbOCH$_3$ and YbOCD$_3$. Calculated values are in the harmonic approximation. Previously measured values for alkaline-earth monomethoxides are included for comparison. All frequencies are in cm$^{-1}$. Typical measurement error bars are $\pm$5~\wavenumber{} depending upon signal-to-noise ratio and proximity to argon emission calibration lines.}
\begin{tabular}{ccccccc}
\hline \hline
Mode & YbOCH$_3$ (meas.) & YbOCH$_3$ (calc.) & YbOCD$_3$ (meas.) & YbOCD$_3$ (calc.) & SrOCH$_3$ $^{a}$& CaOCH$_3$ $^{b}$  \\
\hline

 $\nu_3$ &   1151 & 1154 & - & 1085 & 1138 & 1156 \\ 
 $\nu_4$ &  400 & 390 & 370 & 369 & 405 & 487   \\
 $\nu_8$ & 130 & 134 & 120 & 124 & 135 &   142\\
 $2\nu_8$ & 260 & 268 & - & 248 & - &   -\\
 $\nu_4 + \nu_8$ &  533 & 524 & 500 & 493 &- & 625 \\
 $2\nu_4$ &  805 & 780 & - & 738 & 806&973 \\
 $2\nu_4 + \nu_8$ &  933 & 914 & 880 & 862 & -&- \\
 $\nu_4 + 2\nu_8$ & - & 658 & 680 & 617 &- &- \\
 $\nu_3 + \nu_8$ & 1287 &  1288 & 1270& 1262 & -&-\\
 $3\nu_4$ & 1197 & 1170 & - & - & - & -\\
 $4\nu_4$ & 1576 & 1560 & - & - & - & -\\
 $5\nu_4$ & 1972 & 1950 & - & - & - & -\\

\hline \hline
\multicolumn{5}{l}{$^{a}$ Measured values from Refs.~\cite{OBrien1988} and \cite{brazier1986laser}}\\
\multicolumn{5}{l}{$^{b}$ Measured values from Ref.~\cite{brazier1986laser}}\\
\end{tabular}
\label{tab:VibFreqs}
\end{centering} 
\end{table*}

\subsection{CW Excitation and DLIF Spectra}
\begin{figure*}[t]
\centering
\includegraphics[width=2\columnwidth]{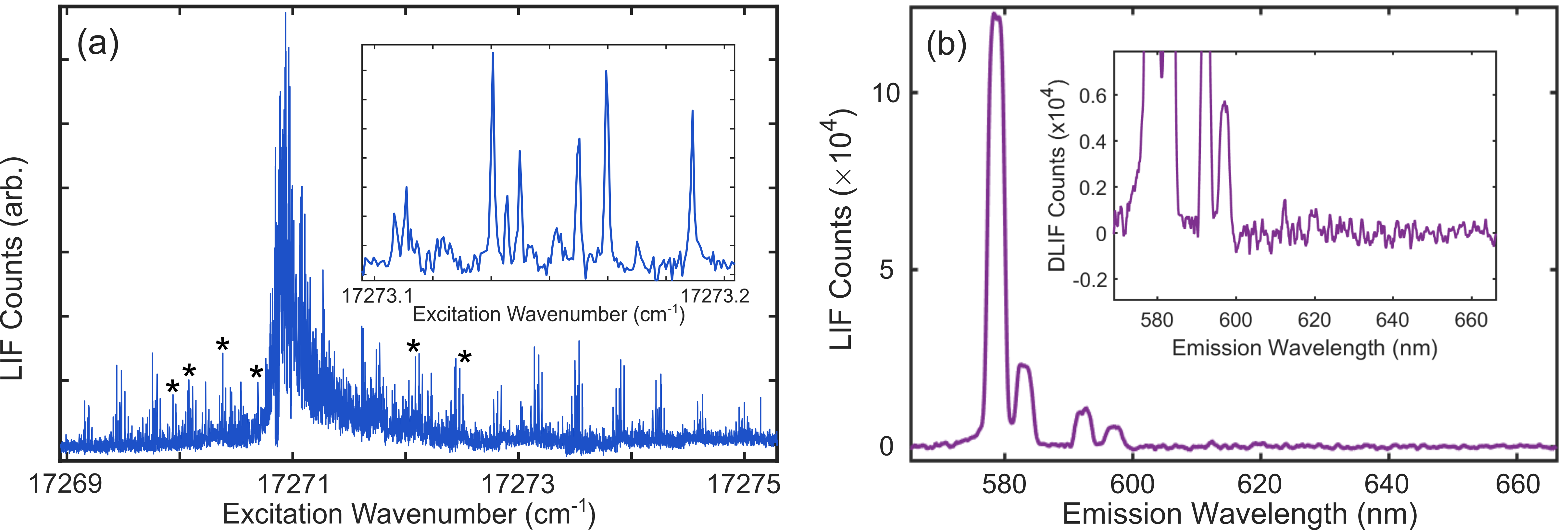} 
\caption{(a) High-resolution excitation spectrum near the YbOCH$_3$ $0_0^0 \, \tilde{A}\,^2E_{1/2} \leftarrow \tilde{X} \, ^2 A_1$ bandhead. Features marked by an asterisk were selected for cw-excitation DLIF measurements. Inset: Details of spectrum near 17273.15~\wavenumber{}, with isotopic and $K$-stack structure visible. (b) DILF spectrum resulting from rotationally-resolved, cw-dye laser excitation at 17270.970~\wavenumber{}. Inset: Zoom near signal baseline to show the relative intensity sensitivity of the measurement.}
\label{fig:HiRes}
\end{figure*}

The high-resolution spectrum recorded near the $0^0_0 \, \Astate_{1/2} \leftarrow \Xstate$ bandhead identified above is shown in Fig.~\ref{fig:HiRes}(a). The spectrum is quite congested due to the presence of several isotopes of Yb with relatively high natural abundance, including $^{174}$Yb (32\%), $^{172}$Yb (22\%), $^{173}$Yb (16\%), $^{171}$Yb (14\%), and $^{176}$Yb (13\%). A full analysis of the high-resolution spectra will be the focus of a future publication, where lower rotational temperature and isotope-selective enhanced production~\cite{Bernath1997, Jadbabaie2020} is used to greatly reduce the complexity of the spectrum. We make some qualitative observations here, showing the spectrum is consistent with a $^2E_{1/2} \leftarrow ^2A_1$ transition.

The rotational energy level structure of a $^2A_1$ state is that of an open-shell prolate symmetric top molecule near the Hund's case (b) limit. For an MOCH$_3$ species, the $K_R^{\dprime}=1$ stack of levels begins approximately 6~\wavenumber{} ($\approx A^{\dprime}$) above the $K_R^{\dprime}=0$ stack. The $\lvert K_R^{\dprime} \rvert =1$ levels obey different nuclear spin statistics and will also be populated significantly and show prominently in the spectrum. The energy level pattern of a $^2E_{1/2}$ state is that of an open-shell prolate symmetric top near the Hund's case (a) limit. The parity doubling in the $K_R^\prime=0, K^\prime=1$ level is expected to be large and hence the dominant $\Astate_{1/2}(K_R^\prime=0,K^\prime=1) \leftarrow \tilde{X} \,^2A_1(K_R^{\dprime}=0)$ band is expected to have the same general appearance as the $\tilde{A}\,^2\Pi_{1/2} \leftarrow \tilde{X}\,^2\Sigma^+$ band of YbOH. Indeed, we observe that the high-resolution spectrum is reminiscent of the $\tilde{A} \, ^2\Pi_{1/2} \leftarrow \tilde{X} \, ^2\Sigma^+$ origin band of YbOH due to the sharply peaked, blue-degraded bandhead and the widely spaced peaks in the $P$/$R$ branches~\cite{Steimle2019, Mengesha2020}. Using the expected YbOCH$_3$ rotational constant, $B$, determined by either empirically scaling parameters for YbOH or by \textit{ab initio} calculations, we estimate that these peaks are spaced by approximately $4B$. This branch spacing implies an excited state parity doubling $\epsilon_1 - 2h_1 \approx B$~\cite{HerzbergVol1}.\footnote{In this expression, $\epsilon_1$ is a higher-order spin-rotation parameter and $h_1$ is a Jahn-Teller parameter~\cite{Crozet2005}.} Here, $\epsilon_1 - 2h_1$ plays the role of the usual $p+2q$ ($\Lambda$-type doubling) parameter in a linear $^2\Pi$ state~\cite{Dick2006SrCH3}. Note that a similar branch spacing of about $4B$ is observed in the YbOH $\tilde{A} \, ^2\Pi_{1/2} \leftarrow \tilde{X} \, ^2\Sigma^{+}$ origin band~\cite{Steimle2019}.

We recorded several DLIF spectra using single-frequency, cw-dye laser excitation on selected rotational lines. These measurements were used to eliminate the possibility that the strong symmetry-forbidden decays observed above originate from the pulsed laser simultaneously exciting many quantum states and/or isotopologues. We selected a few rotationally resolved transitions for cw-excitation DLIF spectra. The selected transitions, marked by asterisks in Fig.~\ref{fig:HiRes}(a), were chosen based on their relative isolation and preliminary assignment of $K^{\prime}=0$ and $K^{\prime}=1$ character. A representative cw-excitation DLIF spectrum is shown in Fig.~\ref{fig:HiRes}(b). The sensitivity of the cw-dye laser measurements is lower than that of the pulsed dye laser measurements due to decreased signal and an inability to temporally exclude the excitation light. This loss in sensitivity required enlarging the entrance slit on the monochromator, lowering the spatial resolution. DLIF spectra recorded on other rotationally-resolved transitions yielded VBRs consistent with those of Fig.~\ref{fig:HiRes}(b) and with the pulsed-laser excitation spectra. These measurements confirmed that the VBRs determined from the pulsed-dye laser excitation are not strongly dependent on the rotational state selected.

\section{Analysis}
\subsection{Vibrational Structure}
The wavelength axes of the DLIF spectra were calibrated and linearized prior to data collection. Ground-state vibrational intervals were determined from the DLIF spectra by locally fitting each peak and taking appropriate differences. The measured frequencies and assignments are listed in Table~\ref{tab:VibFreqs} along with comparisons to the alkaline-earth monomethoxides SrOCH$_3$ and CaOCH$_3$. The observed ground state vibrational energies were fit to the phenomenological expression~\cite{HerzbergVol3}

\begin{equation}
    E(v) = \sum_{i} \omega_i \left(v_i + \frac{d_i}{2} \right) + \sum_{i,k} x_{ik} \left(v_i + \frac{d_i}{2} \right)^2,
    \label{eq:VibFreqs}
\end{equation}

\noindent where mode $i$ has degeneracy $d_i$, frequency $\omega_i$, and $v_i$ quanta of excitation. The constants $x_{ik}$ denote the leading anharmonic corrections. Only the observed modes ($\nu_3$, $\nu_4$, and $\nu_8$)  were included in the fit. The fit incorporated DLIF traces with pulsed-dye laser excitation to the $\tilde{A}(v=0)$, [17.64], and [17.68] states. The results of this fit are presented in Sec.~\ref{sec:DiscXState}.

\subsection{Vibrational Branching Ratios}
Vibrational branching ratios (VBRs), $b_{v^\prime, v^{\dprime}}$, were determined by the ratios of integrated areas under each observed peak. These are related to the FCFs by~\cite{nguyen2018fluorescence}

\begin{equation}
    \begin{split}
    b_{v^\prime,v^{\dprime}} &= \frac{I_{iv^\prime,fv^{\dprime}}}{\sum_{fv^{\dprime}} I_{iv^\prime,fv^{\dprime}}} \\ & \approx \frac{\text{FCF}_{iv^\prime,fv^{\dprime}} \times \nu^3_{iv^\prime,fv^{\dprime}}}{\sum_{fv^{\dprime}}\text{FCF}_{iv^\prime,fv^{\dprime}} \times \nu^3_{iv^\prime,fv^{\dprime}}},
    \end{split}
\label{eq:VBRFCF}
\end{equation}

\noindent where $i$ and $f$ label initial and final states, respectively. Table~\ref{tab:VBRs} lists the observed vibrational branching ratios and FCFs determined for the YbOCH$_3$ $\Astate_{1/2}-\Xstate$, [17.64]$-\Xstate$, and [17.68]$-\Xstate$ bands. The spectrometer and ICCD relative intensity sensitivities were calibrated prior to data collected. The typical noise level of the DLIF spectra corresponds to $\sim$100 counts per pixel, equivalent to $\sim$1 part in $10^3$ relative to the dominant peak. Systematic errors associated with the calibration lead to $\sim$5\% relative uncertainty in peak height near 580~nm and $\sim$10\% relative uncertainty near 630~nm.

\subsection{Electronic Structure Calculations}
To aid assignment of the DLIF spectra, \textit{ab initio} calculations were performed for YbOCH$_3$ and YbOCD$_3$ using the ORCA quantum chemistry program~\cite{neese2012orca, neese2020orca}. Molecular orbitals, optimized geometries, normal modes, and vibrational frequencies for the ground electronic state were calculated at the level of unrestricted Kohn-Sham (UKS) density-functional theory (DFT) using the B3LYP functional~\cite{Bykov2015, Stephens1994}. We carefully studied the accuracy of various basis set and functional choices. Final calculations employed a quadruple-$\zeta$ basis set, including two sets of polarization functions and diffuse orbitals on the heavy atoms. A 28-electron small-core pseudopotential was used for the Yb atom~\cite{Dolg1989, ORourke2019, isaev2015polyatomic}. Time-dependent DFT (TD-DFT) was used to model the $\Astate$ state, using the same basis and functional as were used for the ground electronic state. This allowed computation of excitation energies and geometry optimization using analytic Hessians. Vibrational frequencies were determined via numerical differentiation. The theoretical methods were validated by performing test calculations for YbOH and CaOCH$_3$. See App.~\ref{app:AbInitio} for comparison between different basis sets, validation against previous measurements, and additional computational outputs.

\begin{table*}[tb]
\setlength{\tabcolsep}{9pt}
\begin{centering}
\renewcommand{\arraystretch}{1}
\caption{Branching ratios and FCFs from the $\tilde{A}\,^2E_{1/2} (v=0)$, [17.64], and [17.68] states of YbOCH$_3$ to the $\tilde{X}\,^2A_1$ levels. Values are determined from the pulsed-dye laser excitation DLIF spectra. After accounting for the detection noise and systematic uncertainties due to calibration, the noise levels of the smallest VBRs are $\pm0.7\times10^{-3}$ for the $0^0_0$ band and $\pm2\times10^{-3}$ for the [17.64] and [17.68] bands.}
\begin{tabular}{ccccccc}
\hline \hline
 & \multicolumn{2}{c}{$\tilde{A}\,^2E_{1/2}(v=0)$} & \multicolumn{2}{c}{[17.64]} & \multicolumn{2}{c}{[17.68]} \\
 & VBR & FCF & VBR & FCF & VBR & FCF \\ \hline
$0_0$ & 0.7184 & 0.729 & 0.132 & 0.122 & 0.526 & 0.504 \\
$8_1$ & 0.1653 & 0.164 & - & - & - & - \\
$4_1$ & 0.0566 & 0.054 & 0.624 & 0.617 & 0.375 & 0.385\\
$4_1 8_1$ & 0.0413 & 0.038 & 0.0181 & 0.0183 & 0.044 & 0.046 \\
$4_2$ & $1.1\times10^{-3}$ & $9.89\times 10^{-4}$ & 0.201 & 0.214 & 0.0268 & 0.0293 \\
$4_2 8_1$ & $7.9\times10^{-3}$ & $6.8 \times 10^{-3}$ & - & - & - & - \\
$3_1$ & $6.7\times10^{-3}$ & $5.49 \times 10^{-3}$ & - & - & - & - \\
$3_1 8_1$ & $2.65\times10^{-3}$ & $2.13 \times 10^{-3}$ & - & - & - & - \\
$4_3$ & - & - & 0.021 & 0.0244 & $5.13\times10^{-3}$ & $6.02 \times 10^{-3}$\\
$4_4$ & - & - & $3.6\times10^{-3}$ & $4.36 \times 10^{-3}$ & 0.015 & 0.0194 \\
$4_5$ & - & - & - &                          & $7.6\times10^{-3}$ & $1.0 \times 10^{-2}$ \\

\hline \hline

\end{tabular}
\label{tab:VBRs}
\end{centering} 
\end{table*}

\section{Discussion}
The primary purpose of this investigation was to observe YbOCH$_3$ and determine its suitability for laser cooling and precision tests of fundamental physics. The short radiative lifetime and reasonably diagonal FCFs indicate that YbOCH$_3$ is a promising candidate for these applications, similar to YbF~\cite{Sauer1996,Sauer1999,Lim2017} and YbOH~\cite{Nakhate2019}. The data provide information necessary to begin further studies of optical cycling and vibrational repumping. As discussed below, YbOCH$_3$ also presents an interesting platform for future studies of (pseudo-)Jahn-Teller interaction in the presence of strong spin-orbit coupling.

\subsection{The $\Xstate$ State}
\label{sec:DiscXState}
Assignments of the observed vibrational decays were relatively straightforward using the \textit{ab initio} predictions, isotopic data, and comparison to alkaline-earth monomethoxides. For the origin band, the peaks redshifted by 130~\wavenumber{} and 400~\wavenumber{} are assigned to  $8^0_1$ and $4^0_1$, respectively. The feature redshifted by 1150~\wavenumber{} is near the expected frequency of either $\nu_3$ (1$\times$ O-C stretch) or $3\nu_4$ (3$\times$ Yb-O stretch). Because this decay has a stronger intensity than $4^0_2$, we favor the assignment to $3^0_1$. Furthermore, the DLIF spectra from the [17.64] and [17.68] states show a feature at slightly higher frequency ($\approx 1170$~\wavenumber{}) that is assigned to $3\nu_4$. The weak feature redshifted by $\sim$1280~\wavenumber{} is assigned to the decay $3^0_1 8^0_1$ based on the predicted frequencies for various combination bands. In addition, the relative intensity of this feature to the $3^0_1$ peak is roughly consistent with the strengths of other combination bands involving $\nu_8$.

The vibrational frequencies are determined by fitting observed vibrational energy intervals to Eq.~\ref{eq:VibFreqs}. We find $\omega_3 = 1152(5)$~\wavenumber{}, $\omega_4 = 405(3)$~\wavenumber{}, $\omega_8 = 130(3)$~\wavenumber{}, and $x_{44} = -2.0(6)$~\wavenumber{}. These fitted frequencies agree well with the computed harmonic frequencies and with the corresponding values for CaOCH$_3$ and SrOCH$_3$ after accounting for differences in reduced mass~\cite{brazier1986laser}. The anharmonic contribution to the Yb-O stretching mode is similar to that observed in YbOH in both magnitude and sign~\cite{Mengesha2020}. Finally, the fitted value $\omega_3$ is similar to that predicted for the methoxy anion, consistent with a highly ionic metal-ligand bond~\cite{Weil1985}.

Our \textit{ab initio} calculations predict that the $\Xstate$ state has a linear Yb-O-C bond and symmetric, off-axis H atoms with $C_{3v}$ symmetry, just like the alkaline-earth monomethoxides. The calculated values $r_{\rm OC} = 1.39$~\AA~and $r_{\rm CH} = 1.10$~\AA~for YbOCH$_3$ are quite similar to those measured in CaOCH$_3$ ($r_{\rm OC} = 1.411(7)$~\AA~and $r_{\rm CH}=1.094$~\AA)~\cite{Crozet2005}, while the calculated value of $r_{\rm YbO} = 2.04$~\AA~for YbOCH$_3$ is very close to the measured value for YbOH ($r_{\rm YbO} = 2.0397$~\AA)~\cite{Nakhate2019}. In addition, our \textit{ab initio} calculations predict a molecule-frame dipole moment of $\mu(\tilde{X}) = 2.2$~D in the $\Xstate$ ground state. This is nearly identical to the ground-state dipole moment of YbOH (1.9(2)~D~\cite{Steimle2019}) and comparable to that of CaOCH$_3$ (1.58(8)~D~\cite{Namiki1998}). These calculations also report the $\Xstate$ electronic state arises primarily from Yb-centered $6s\sigma$ (65\%) and $6p\sigma$ (15\%) orbitals. This is quite similar to YbF where experimental analysis of the hyperfine structure showed that the $X \, ^2\Sigma^+$ state had 57\% $6s$ character~\cite{Steimle2007}. The resulting polarization of the valence electron away from the Yb-O bond is consistent with the predicted molecule-frame dipole moment being significantly smaller than that of an electrostatic model with point charges located near the Yb$^{+}$ and OCH$_3^{-}$ moieties~\cite{Marr1996}. Together, these results help confirm the expectation that YbOCH$_3$ in the $\Xstate$ ground state conforms to many of the patterns established by previous studies of alkaline-earth monomethoxides. The predicted vibrational frequencies for YbOCH$_3$ and YbOCD$_3$ show excellent agreement with our measurements, as can be seen in Tab.~\ref{tab:VibFreqs}. Although we only observed activity in three of the vibrational modes, we report the full results of the vibrational structure in App.~\ref{app:AbInitio}.

\subsection{The Excited States}
\label{sec:DiscAState}
We have observed a number of excited states assigned to YbOCH$_3$, associated with bands near 17271~\wavenumber{}, 17281~\wavenumber{}, 17641~\wavenumber{}, 17683~\wavenumber{}, and 18542~\wavenumber{}. The band at 17271~\wavenumber{} is assigned to terminate on the $\Astate_{1/2} (v=0)$ level. As discussed above, both the [17.64] and [17.68] states seem to have strong $v_4=1$ (Yb-O stretching) character, but on the basis of the FCF measurements we preliminarily assign the [17.64] state as dominantly $\Astate_{1/2}(v_4=1)$, while the [17.68] state appears to be a perturbing state similar to those observed in both YbF~\cite{Lim2017} and YbOH~\cite{Mengesha2020}. The band near 18542~\wavenumber{} is tentatively assigned as $0^0_0 \, \tilde{A}\,^2E_{3/2} \leftarrow \Xstate$. This implies an effective spin-orbit splitting $a \zeta_e d \approx 1270$~\wavenumber{}, similar to that of YbF~\cite{Dunfield1995} and YbOH~\cite{SteimleTBD}. Future high-resolution studies will be useful to provide further confirmation of these assignments.

Observation of the relatively intense $8^0_1$ feature in the $\tilde{A}\,^2E_{1/2}(v=0)$ DLIF spectrum (Fig.~\ref{fig:DLIF}) was at first unexpected because, assuming both the $\tilde{X}$ and $\tilde{A}$ states have $C_{3v}$ symmetry, the $\nu_8$  mode is of $e$ symmetry and this decay is symmetry-forbidden within the Born-Oppenheimer (BO) approximation. One possible explanation for the strength of this transition is that YbOCH$_3$ possesses a bent geometry ($C_s$ symmetry) in the $\tilde{A}$ state. We disfavor this explanation for several reasons. First, we do \textit{not} observe emission from $\Astate_{1/2}$ to $\Xstate (v_8=2)$, which would be expected if there were significant geometrical change upon electronic excitation. Similarly, we do not observe strong decays to $\Xstate (v_8=1)$ from the [17.64] or [17.68] states. Second, a bent excited state would likely be accompanied by significant quenching of the electronic orbital angular momentum, but the above assignment of the $\tilde{A}\,^2E_{3/2}$ state is consistent with $a \zeta_e d \approx 1270$~\wavenumber{}, indicating little quenching. Third, we did not observe rotational bands associated with the $K_a$ structure expected for a bent molecule; the contour of the high-resolution spectra could be well simulated using a symmetric top Hamiltonian. A more detailed discussion of the high-resolution spectrum will be the subject of a separate publication.

Instead, we attribute the relatively strong $8^0_1$ signal to spin-orbit vibronic coupling. Following Ref.~\cite{Paul2019}, we consider the possibility of Jahn-Teller coupling within $\Astate$, characterized by parameter $k_8$, and pseudo-Jahn-Teller coupling between $\Astate$ and nearby states of $A_1$ symmetry, characterized by parameter $\lambda_8$. Here, we assume that $\nu_8$ is the only mode with significant Jahn-Teller activity. Spin-orbit coupling between $\Astate$ and $\tilde{B}\,^2A_1$ can interfere with (pseudo-)Jahn-Teller terms and lead to a second-order, effective coupling between the $\Astate(v_8=1)$ and $\Astate(v_8=0)$ vibronic levels. Given the intensity of the $8^0_1$ feature, we infer an effective matrix element $$\langle \Astate_{1/2},v_8=1 \rvert H_{JT} \lvert \Astate_{1/2},v_8=0 \rangle \approx 50~\text{\wavenumber{}}.$$ Again following Ref.~\cite{Paul2019}, we may also express this matrix element as $k_8 + 2C_{AB}\lambda_8$. The first term represents direct coupling between $\Astate(v_8=0,1)$ while the second term represents the spin-orbit vibronic coupling via an intermediate $^2A_1$ state. $C_{AB}$ is the mixing induced by spin-orbit coupling between $\Astate$ and $\tilde{B}\,^2A_1$, approximated using the measured spin-orbit parameter of YbOCH$_3$ and the estimated $\tilde{B}\,^2A_1$ energy scaled from YbF. It has been shown for CaOCH$_3$ that the $k_8$ term is negligible compared to the $\lambda_8$ term. Assuming the same holds for YbOCH$_3$, we estimate $\lambda_8 \approx 55$~\wavenumber{}. This is approximately three times smaller than the value of $\lambda_8$ computed by \textit{ab initio} methods for CaOCH$_3$, which is consistent with the expected trend as spin-orbit coupling strength increases~\cite{Barckholtz1998}. Though preliminary, this estimate shows that the strong $8^0_1$ feature can be explained reasonably by the spin-orbit vibronic coupling mechanism. 

Recent studies of CaOCH$_3$ by our group lend support to these quantitative estimates. In those measurements, which will be discussed in more detail elsewhere, we observed the analogous vibronically-induced $8^0_1$ feature in CaOCH$_3$, albeit with weaker VBR $\approx 3 \times 10^{-3}$. Note that this agrees quite well with the \textit{ab initio} prediction of Ref.~\cite{Paul2019}, where the decay was predicted but not observed. Applying the same analysis as used above, we determine a linear pseudo-Jahn-Teller coefficient for CaOCH$_3$ of $\lambda_{8}^\text{CaOCH$_3$} \approx 145$~\wavenumber{}. This is in good agreement with the \textit{ab initio} value of $150$~\wavenumber{}~\cite{Paul2019} and provides a satisfactory check of the estimation methods used above. Nonetheless, \textit{ab initio} treatment of the vibronic coupling problem in YbOCH$_3$ would be highly desirable. In addition, we note that other interpretations are possibly valid, such as the $k_8$ parameter being nonnegligible or vibronic coupling between $\Astate$ and nearby electronic states arising from excitations of Yb $4f$ electrons. The presence of a strong perturbation makes it very challenging to use standard semi-empirical methods~\cite{Mengesha2020,kozyryevCaOCH3, Augenbraun2020ATM} to predict the FCFs in YbOCH$_3$. This reinforces the need for careful experimental studies to guide future experiments with this molecule.

TD-DFT calculations targeting the $\Astate$ state predict an excitation energy $T_e = 18060$~\wavenumber{}. This is in good agreement with the measured value of $T_e \approx 17900$~\wavenumber{} (accounting for the spin-orbit coupling). The $\Astate$ state is predicted to have $C_{3v}$ symmetry, with bond lengths/angles $r_{\rm YbO} = 2.010$~\AA, $r_{\rm OC} = 1.396$~\AA, $r_{\rm CH} = 1.095$~\AA, and $\theta_\text{OCH} = 111.283^\circ$. The calculations predict a linear Yb-O-C structure. The predicted change in Yb-O bond length is quite similar to that observed for YbOH and consistent with the relatively diagonal branching ratios~\cite{Steimle2019}. The calculation predicts a vibrational frequency $\omega_4 = 410$~\wavenumber{}, which is in reasonable agreement with the splitting observed between the $\Astate$ and [17.64]/[17.68] states. See App.~\ref{app:AbInitio} for a full set of predicted excited-state vibrational frequencies. 

\subsection{Prospects for Laser Cooling and Trapping}

Our measurements confirm the viability of achieving rapid photon cycling with YbOCH$_3$. The short radiative lifetime of the $\tilde{A}$ state (corresponding to $\Gamma \approx  27 \times 10^6$~s$^{-1}$) will enable large optical forces, comparable to those of previously laser-cooled molecules. Based on our VBR measurements, a feasible laser cooling scheme requiring about 8 distinct laser wavelengths will achieve vibrational closure below a part in $10^3$.  Additional frequencies required for repumping rotational and spin-rotation substructure can be added as laser frequency sidebands using electro- or acousto-optic modulators. Assuming a cryogenic buffer-gas beam containing $10^9$ YbOCH$_3$ molecules and a loss probability per photon scatter of $<1 \times 10^{-3}$ (the smallest observed VBR) to unaddressed vibrational states, this scheme would leave $\sim 10^4 - 10^5$ YbOCH$_3$ molecules in a ``bright" state following $\sim$10,000 photon scatters. This is comparable to the experimental complexity required to laser cool much lighter (and simpler) species~\cite{Baum2020PRA, Mitra2020, kozyryevCaOCH3}. Laser-enhanced molecule production such as that demonstrated for YbOH~\cite{Jadbabaie2020} could increase the molecule number by at least an order of magnitude. The presence of multiple excited electronic states with strong electric-dipole allowed transitions to excited vibrational states in $\tilde{X}$ provides a convenient method of decoupling the repumping pathways, similar to the scheme used for YbF~\cite{Lim2018}.

\subsection{Suitability for Precision Measurements}
\label{sec:DiscPrecisionMeasurements}
To the best of our knowledge, YbOCH$_3$ is the highest-$Z$ symmetric top molecule amenable to optical cycling to be experimentally characterized. It is a promising system for precision tests of parity and/or time-reversal symmetry violation. In addition to the high-$Z$ nucleus that leads to a valence electron experiencing strong relativistic effects, the presence of easily polarized, internal co-magnetometer states in the $\lvert K^{\dprime} \rvert \neq 0$ manifolds allow full access to a large internal effective electric field and rejection of potential systematic errors~\cite{Kozyryev2017PolyEDM, Hutzler2020}. Consider, for example, laser-cooled and trapped YbOCH$_3$ used for future tests of the eEDM. Under the reasonable assumptions that the internal effective electric field of YbOCH$_3$ matches that of YbOH ($\mathcal{E}_\text{eff} \approx 25$~GV/cm)~\cite{Denis2019, Gaul2019, Prasannaa2019} and that achievable experimental lifetimes approach $10-100$~s~\cite{OHara1999, Bernon2013, Wang2018}, $10^5$ trapped YbOCH$_3$ molecules and one week of averaging could provide a statistical sensitivity to the eEDM four orders of magnitude beyond the current limit~\cite{ACME2018}. This is about an order of magnitude more sensitive than an equivalent measurement in YbOH.

Even with a much smaller photon budget, transverse laser cooling and high-fidelity readout enabled by rapid photon cycling could lead to new limits on precision tests of $CP$-violation~\cite{Ho2020}. Consider an experiment using $^{173}$YbOCH$_3$ in a molecular beam to search for a nMQM. Combining the demonstrated production of YbOCH$_3$, enhanced chemical production~\cite{Jadbabaie2020}, and transverse laser cooling using $<$150 photons~\cite{Mitra2020}, bright beams with $>10^7$ molecules per pulse in the science state appear possible. Forward velocities around 50~m/s and laser cooling to $< 1$~mK in the transverse direction~\cite{Augenbraun2020YbOH, Mitra2020} would make coherence times on the order of 10~ms feasible. Under these conditions, a sensitivity on the order of 10~$\mu$Hz would be achievable with one week of averaging. Assuming (reasonably) that the nMQM sensitivity parameter is similar to that of YbOH, this would be near or below the level required to set new limits on $T$-violating BSM physics, e.g., the QCD $\bar{\theta}$ parameter or the difference of quark chromo-EDMs~\cite{Maison2019, Denis2020MQM}. Note that symmetric top molecules offer some intrinsic advantages over linear species for such measurements such as requiring smaller laboratory electric fields for full polarization and that nuclear spin statistics ease the task of state preparation by naturally populating the $\lvert K^{\dprime}\rvert=1$ science state~\cite{Kozyryev2017PolyEDM, Yu2020}.

YbOCH$_3$ is also an interesting candidate for possible tests of $P$-violation. Recently, the degenerate bending modes in open-shell linear triatomic molecules have been proposed for tests of nuclear spin-dependent parity violation~\cite{Norrgard2019, Hao2020}. Similar searches could be conducted in symmetric top molecules that meet certain structural conditions. One key requirement of such an experiment is that the ground-state parity doubling is not smaller than all spin-rotation/hyperfine structure, such that opposite-parity states with the same total angular momentum may be Zeeman-tuned into near resonance. In $^{171}$YbOCH$_3$, we expect that hyperfine and $K$-doublet splittings will be of the same order of magnitude, $\sim$0.3~MHz~\cite{Namiki1998}. The size of the $K$-doubling increases with $N$, so that by working in, e.g., the $N^{\dprime}=2$ or $3$ state it may be possible to tune the parity-doubling to be larger than the hyperfine structure. Future measurements with hyperfine resolution will be required to determine the exact splittings. Alternatively, chiral variants such as YbOCHDT may be useful for studies of parity violation arising from $P$-odd cosmic fields, e.g., in searches for axion-like particles~\cite{Gaul2020a, Gaul2020b}. Chiral species are predicted to be particularly sensitive to oscillating pseudovector fields that cannot be easily probed in existing spin-precession experiments~\cite{Gaul2020a}. Because such effects scale rapidly with $Z$~\cite{Gaul2020b}, YbOCHDT may provide orders-of-magnitude increased sensitivity over previously considered chiral probes. Production and characterization of chiral ytterbium monomethyl, YbCHDT, may be especially promising by bringing the heavy nucleus closer to the chiral center.

\section{Conclusion}
We have observed ytterbium monomethoxide, YbOCH$_3$, and performed initial measurements to assess its feasibility for laser cooling and precision measurements of $CP$-violating electromagnetic moments. The measured vibrational frequencies, vibrational branching ratios, and radiative lifetimes are compatible with a direct laser cooling experiment. Electronic structure calculations have aided the vibrational assignments. We have interpreted some of the vibrational branching ratios using a simple linear vibronic coupling model. Despite having stronger off-diagonal decays than the isoelectronic species YbF and YbOH, the vibrational branching ratios in YbOCH$_3$ converge rapidly enough to permit efficient laser cooling. Even in the absence of full 3D laser cooling, YbOCH$_3$ would allow optical cycling of $\sim$150 photons with about 3 vibrational repumping lasers. This scale enables efficient state preparation, transverse cooling, and/or unit-efficiency readout for precision measurements in a molecular beam.

Further data are needed to sharpen the developing picture of the $\Astate_{1/2} - \Xstate$ band. The $\tilde{A} \, ^2\Pi$ states in YbF and YbOH are quite complex due to a series of perturbing states nearby in energy~\cite{Mengesha2020, Lim2017}, and it appears that analogous states exist in YbOCH$_3$. Careful studies of the $\tilde{A}$ state and nearby perturbing levels will be necessary to develop a laser cooling scheme capable of scattering $\gg 10^3$ photons with YbOCH$_3$. Detailed understanding of vibronic interactions will also inform the feasibility of laser cooling increasingly complex polyatomic molecules~\cite{Dickerson2020}. Finally, vibronic interactions with matrix elements off-diagonal in $K$ may lead to leakage from the optical cycle at rates high enough to require repumping. Characterizing these effects using high-resolution spectroscopy will also be important prior to laser cooling YbOCH$_3$.

\section{Acknowledgments}
We are grateful to Nick Hutzler, Ivan Kozyryev, and Phelan Yu for constructive feedback on this manuscript. We thank Lan Cheng and the PolyEDM collaboration for many useful discussions. This work was supported by the W. M. Keck Foundation and the Heising-Simons Foundation. BLA acknowledges financial support from the NSF GRFP. ZDL was supported by the Center for Fundamental Physics (Fundamental Physics Grant) and the Templeton Foundation. HS was supported by the Ezoe Memorial Recruit Foundation. The computations in this paper were run on the FASRC Cannon cluster supported by the FAS Division of Science Research Computing Group at Harvard University.

\appendix

\section{Additional information on \textit{ab initio} calculations}
\label{app:AbInitio}
Here, we provide additional data to support our electronic structure calculations. Our \textit{ab initio} calculations were performed using the ORCA quantum chemistry package~\cite{neese2012orca}. Ground state vibrational frequencies and transition dipole moments were computed using analytic B3LYP Hessians~\cite{Stephens1994,Bykov2015}. The excited state was modeled using time-dependent density functional theory (TD-DFT) using the same functional. The excited state vibrational structure was computed by numerical differentiation. Initial studies used the def2-TZVPP basis set, while final calculations employed a ma-def2-QZVPP basis set~\cite{Zheng2010}. A 28-electron small-core pseudopotential was used for the Yb atom~\cite{Dolg1989, ORourke2019, isaev2015polyatomic}. The addition of diffuse functions to the basis set improved agreement between theory and experiment for both dipole moment and vibrational frequencies. A comparison between these choices of basis sets can be found in Tabs.~\ref{tab:YbOHAbInitio} and \ref{tab:CaOCH3AbInitio}. To validate our theoretical methods, we performed additional calculations for previously observed, relevant species. Here, we report the comparison of these test calculations with experimental data. We also show the effect of basis set quality on the computed values. 

Table~\ref{tab:YbOHAbInitio} reports several predictions for the $\tilde{X}\,^2\Sigma^+$ and $\tilde{A}\,^2\Pi$ electronic states of YbOH, including vibrational frequencies, geometry, and molecule-frame electric dipole moment. For the ground state properties, we find excellent agreement between theory and experiment. We show the predicted vibrational frequencies, bond lengths, and excitation energy of the excited state. The agreement is satisfactory for the excited state properties, with errors similar to that observed in \textit{ab initio} calculations for related species~\cite{Paul2019, nguyen2018fluorescence}. The agreement in these cases supports the choice of method, especially with regard to modeling the heavy Yb atom.

Table~\ref{tab:CaOCH3AbInitio} reports predictions for vibrational frequencies, geometry, and molecule-frame dipole moment of the $\Xstate$ electronic state of CaOCH$_3$. Again, we observe excellent agreement with experimental measurements. These calculations help to validate the accuracy of our computational methods for related alkaline-earth monomethoxide species.

In the experiment, only three of the eight vibrational modes were observed. Table~\ref{tab:YbOCH3VibFull} reports all of the predicted vibrational frequencies for YbOCH$_3$, obtained using the quadruple-$\zeta$ basis set described above. We also include the predicted vibrational frequencies of the $\Astate$ state of YbOCH$_3$.

\begin{table}[tb]
\setlength{\tabcolsep}{9pt}
\begin{centering}
\renewcommand{\arraystretch}{1}
\caption{Calculated properties of YbOH $\tilde{X} \, ^2\Sigma^+$ state and comparison to measured values using different basis sets.}
\begin{tabular}{lccc}
\hline \hline
Parameter & Calc. 1\footnote{Using def2-TZVPP basis.} & Calc. 2\footnote{Using ma-def2-QZVPP basis.} & Meas.~\cite{Steimle2019,Mengesha2020, Nakhate2019} \\
\hline
\multicolumn{4}{l}{Vibrational Frequencies [cm$^{-1}$]} \\
 $\nu_1(\tilde{X})$ & 538 & 530 & 529.33(2) \\
 $\nu_2(\tilde{X})$ & 335 & 332 & 329(14)  \\
 $\nu_3(\tilde{X})$ & 3940 & 3937 & - \\
 $\nu_1(\tilde{A})$ & - & 560 & 584\footnote{Taken as the $\tilde{A}(100) - \tilde{A}(000)$ separation from Ref.~\cite{Steimle2019}.} \\
 $\nu_2(\tilde{A})$ & - & 329 & 320\footnote{Taken as the [17.64]$-\tilde{A}(000)$ separation from Ref.~\cite{Steimle2019}.}  \\
 $\nu_3(\tilde{A})$ & - & 3950 & - \\ \\
\multicolumn{4}{l}{Bond Lengths [$\mathrm{\AA}$]} \\
$r_\text{YbO}(\tilde{X})$ & 2.0386 & 2.0441 & 2.0397 \\
$r_\text{OH}(\tilde{X})$ & 0.9540 & 0.9530 & 0.9270\footnote{Fixed to value for BaOH, as in Ref.~\cite{Nakhate2019}.} \\ 
$r_\text{YbO}(\tilde{A})$ & - & 2.012 & 2.006 \\
$r_\text{OH}(\tilde{A})$ & - & 0.9533 & 0.9270\footnote{Fixed to value for BaOH, as in Ref.~\cite{Nakhate2019}.} \\ \\
\multicolumn{4}{l}{Dipole Moment [D]} \\
$\mu_e(\tilde{X})$ & 1.6 & 1.8 & 1.9(2) \\ \\
\multicolumn{4}{l}{Excitation energy [\wavenumber{}]} \\
$T_e$ & - & 18189 & 17998 \\ \\
\hline \hline
\end{tabular}
\label{tab:YbOHAbInitio}
\end{centering} 
\end{table}

\begin{table}[tb]
\setlength{\tabcolsep}{9pt}
\begin{centering}
\renewcommand{\arraystretch}{1}
\caption{Calculated properties of CaOCH$_3$ $\Xstate$ state and comparison to measured values using different basis sets.}
\begin{tabular}{lccc}
\hline \hline
Parameter & Calc. 1\footnote{Using def2-TZVPP basis.} & Calc. 2\footnote{Using ma-def2-QZVPP basis.}  & Measured~\cite{kozyryevCaOCH3, Crozet2005} \\
\hline
\multicolumn{4}{l}{Vibrational Frequencies [cm$^{-1}$]} \\
 $\nu_3(\tilde{X})$ & 1140 & 1159 & 1156(5) \\ 
 $\nu_4(\tilde{X})$ & 488 & 489 & 488(5) \\
 $\nu_8(\tilde{X})$ & 139 & 141 & 144(5) \\
 
 $\nu_3(\tilde{A})$ & - & 1140 & 1140(5) \\ 
 $\nu_4(\tilde{A})$ & - & 470 & 501.48 \\
 $\nu_8(\tilde{A})$ & - & 150 & 145(5) \\ \\
\multicolumn{4}{l}{Bond Lengths [$\mathrm{\AA}$] and angles [$^\circ$]} \\
$r_\text{CaO}(\tilde{X})$ & 1.958 & 1.969 & 1.962(4) \\
$r_\text{OC}(\tilde{X})$ & 1.400 & 1.392 & 1.411(7) \\
$r_\text{CH}(\tilde{X})$ & 1.105 & 1.097 & 1.0937 \\
$\theta_\text{OCH}(\tilde{X})$ & 111.5 & 111.5 & 111.3(2) \\
\\
\multicolumn{4}{l}{Dipole Moment [D]} \\
$\mu_e(\tilde{X})$ & 1.1 & 1.4 & 1.58(8) \\ \\
\multicolumn{4}{l}{Excitation energy [\wavenumber{}]} \\
$T_e$ & - & 16088 & 15925 \\
\hline \hline
\end{tabular}
\label{tab:CaOCH3AbInitio}
\end{centering} 
\end{table}

\begin{table}[tb]
\setlength{\tabcolsep}{9pt}
\begin{centering}
\renewcommand{\arraystretch}{1}
\caption{Calculated vibrational frequencies for YbOCH$_3$, computed at the optimized geometry, including modes not observed in the LIF spectra presented in this manuscript. See Tab.~\ref{tab:VibLabels} for the mode labeling and symmetry characteristics.}
\begin{tabular}{ccc}
\hline \hline
Vibrational Mode & $\Xstate$ (cm$^{-1}$) & $\Astate$ (cm$^{-1}$) \\
\hline
$\nu_1$ & 2940 & 2961 \\
$\nu_2$  & 1482 & 1523 \\
$\nu_3$ & 1157 & 1156 \\
$\nu_4$  & 390 & 413 \\
$\nu_5$ & 2975 & 3007 \\
$\nu_6$  & 1495 & 1489 \\
$\nu_7$ & 1180 & 1181 \\
$\nu_8$  & 135 & 139 \\
\hline \hline

\end{tabular}
\label{tab:YbOCH3VibFull}
\end{centering} 
\end{table}

\newpage

\bibliography{YbOCH3bib}
\end{document}